\documentclass[aps,prd,onecolumn,eqsecnum,amsmath,nofootinbib,preprintnumbers]{revtex4}%

\usepackage{color,graphicx,float,subfigure}
\usepackage{amsfonts,amssymb,theorem,mathrsfs,times}
\usepackage{bm}
\usepackage{mathtools}
\usepackage{amsfonts,amssymb,theorem,mathrsfs}
\usepackage{dsfont}
\usepackage{setspace}

\usepackage[colorlinks,linkcolor=blue,anchorcolor=blue,citecolor=blue]{hyperref}   
\usepackage{ulem}   

{\theorembodyfont{\upshape}
	}
{\theorembodyfont{\upshape}
	}
{\theorembodyfont{\upshape}
	}
{\theorembodyfont{\upshape}
	}
{\theorembodyfont{\upshape}
	}
{\theorembodyfont{\upshape}
	}
\addtocounter{MaxMatrixCols}{10}
\newcommand{\dalm}{\kern1pt\vbox{\hrule height 0.9pt\hbox{\vrule width
			0.9pt\hskip 2.5pt\vbox{\vskip 5.5pt}\hskip 3pt\vrule width
			0.3pt}\hrule height 0.3pt}\kern1pt}

\begin{document}
\preprint{\hfill {\small {ICTS-USTC/PCFT-23-02}}}
\title{Constraint preserving boundary conditions in Bondi-Sachs gauge: a numerical study of stability of pure AdS spacetime}

%

\author{ Li-Ming Cao$^{a\, ,b}$\footnote{e-mail
		address: caolm@ustc.edu.cn}}

\author{ Liang-Bi Wu$^b$\footnote{e-mail
		address: liangbi@mail.ustc.edu.cn}}

\author{ Yu-Sen Zhou$^b$\footnote{e-mail
		address: zhou\_ys@mail.ustc.edu.cn}}

\affiliation{$^a$Peng Huanwu Center for Fundamental Theory, Hefei, Anhui 230026, China}

\affiliation{${}^b$
	Interdisciplinary Center for Theoretical Study and Department of Modern Physics,\\
	University of Science and Technology of China, Hefei, Anhui 230026,
	China}

\date{\today}

\begin{abstract}
	In the Bondi-Sachs gauge, the Einstein equations with a cosmological constant coupled to a scalar field in spherical symmetry are cast into a first order strongly hyperbolic formulation in which the lapse and shift are the fundamental variables. 
	For this system of equations, the lapse and shift are ingoing characteristic fields, and the scalar field has three modes: ingoing, outgoing and static, respectively. A constraint-preserving initial boundary value problem is constructed by using Bianchi identity. 
	Using this scheme, we find that any small perturbation of the scalar field at the boundary far away enough can  cause the collapse of the pure AdS spacetime, and we provide the numerical evidence for the formation of  apparent horizons.
	The numerical evolution is performed with a standard method of lines, second order in space and time. The evolution is performed using the standard second order Runge-Kutta method while the space discrete derivative is second order central difference with fourth order artificial dissipation.

\end{abstract}


\maketitle

\section{Introduction}\label{section1}
Motivated mainly by the AdS/CFT correspondence, a very basic question ``Is AdS stable?" is raised~\cite{Bizon:2011gg}. In more detail, one may ask ``Under the background with a negative cosmological constant, whether the pure AdS solution will collapse to form a black hole under small perturbations of the scalar field?"
 It is difficult  to solve this problem analytically. However,  one can attack the problem by numerical relativity. Actually, it has been shown in~\cite{Bizon:2011gg} that pure  (global) AdS  spacetime is not stable under the small variation of the initial data.  
Instead of the perturbation on the initial surface,    by using the numerical relativity,   we give some evidence that any small perturbation of the scalar field located on the boundary far away enough can also give rise the collapse of  the pure AdS solution.
Under the perturbation, an apparent horizon will emerge inside the computational domain,  and a black hole forms.
 
 The main task of numerical relativity is to solve Einstein equations numerically under different gauges or coordinates.
There are two common schemes of solving Einstein equations. In one of the schemes, the spacetime is foliated by spacelike hypersurfaces. Each hypersurface represents an instant of time. Proper initial data is described on one time slice. 
By using a version of the Einstein equations, one gets the data on a later time slice. This scheme is the so-called  Cauchy problem for the Einstein equations. The ADM formulation is the original version of the Cauchy problem~\cite{1979Kinematics}. 
Moreover, in order to get a well-posed strongly hyperbolic system, different research groups have converted the ADM formulation to the the Baumgarte–Shapiro–Shibata–Nakamura (BSSN) formulation which is very robust in practice~\cite{Shibata:1995we,Baumgarte:1998te}.

For another scheme which is called the characteristic formulation, spacetime is foliated by null hypersurfaces. Appropriate data are prescribed on an initial retarded time slice and possibly on another hypersurface transverse to the retarded slice.  In numerical relativity, 
the original version of the characteristic problem is the Bondi-Sachs formulation~\cite{Sachs:1962wk}.  Rounded discussion can be found in the review paper by Wincour~\cite{Winicour:2012znc}.

For an actual evolutionary process, there are two alternatives for solving Einstein equations. One is called a constrained scheme. It is a time scheme for integrating the 3+1 Einstein system in which some or all of the four constraints (Hamiltonian and momentum constraints) 
are used to compute some metric coefficients at each step of the numerical evolution. The other is called a free evolution scheme. It is a time scheme for integrating 3+1 Einstein system in which the constraint equations are solved only to get the initial data~\cite{Gourgoulhon:2007ue}.

A straightforward manipulation of the Bianchi identities demonstrates that either strategy produces the same solution at the analytical level~\cite{Calabrese:2001kj}. Hence, there is no need to deal with constrained evolution which usually requires solving elliptic equations. 
Solving elliptic equations at every time step demands a significant computational overhead. For this reason constrained evolutions have been, for the most part, avoided beyond the two dimensional case. The more direct approach of free evolution can be safely employed. 
However, free evolution in numerical implementations display violation of the constraints where only using constraints to get the initial data. Choptuik gives some insights on implementing unconstrained code~\cite{Choptuik:1991bfu}.

When the free evolution scheme is considered, one's purpose is to provide proper boundary conditions so that  there exits a unique solution to the evolution equation, while the constraints are preserved in the entire computational domain. In a strongly hyperbolic formulation of gravity, 
it is sufficient that giving boundary conditions only to the characteristic modes that enter the computational domain at any given boundary~\cite{1995Numerical}. 
Several efforts have illustrated the advantages of providing boundary conditions through the use of constraints in a real numerical simulation~\cite{Iriondo:2001ig}.

Now, in our paper, we are dealing with a first-order, quasi-linear strongly hyperbolic formulation of Einstein equations with a negative cosmological constant. Such a system can be written as
\begin{eqnarray}
	\dot{u}=Mu^{\prime}+\text{l.o.}\, ,\nonumber
\end{eqnarray}
in the spherically symmetry case, where $u$ is a list of variables, the dot and the prime indicate time and spatial derivatives, respectively.  The term l.o.stands for lower order terms which do not have any derivatives. 
$M$ is the so-called  principal part which is a diagonalizable matrix that might depend on the variables  themselves and the spacetime coordinates, but not depend on the derivatives of $u$.  The evolution of the constraints (auxiliary system) is given by
\begin{eqnarray}
	\dot{u}_c=M_cu_c^{\prime}+\text{l.o.}\, .\nonumber
\end{eqnarray}
A unique solution to the auxiliary system is fixed by giving initial data to $u_c$ and boundary conditions to the ingoing modes of the system. Note that characteristic fields (eigenvectors of the principal part $M$) with positive eigenvalues are travelling to the left, on the contrary, characteristic fields (eigenvectors of the principal part $M$) with negative eigenvalues are travelling to the right. Therefore, for the left boundary, the boundary conditions are needed for the negative characteristic field. Accordingly, for the right boundary, the boundary conditions are needed for the positive characteristic field. Since in the case of Einstein equations, the auxiliary system is usually supposed to be a homogeneous system. The identically zero solution is obtained by providing zero as initial data (This is the process of solving the initial data.) and zero boundary conditions to the eigenmodes of $M_c$ entering the domain. More details can found in the Ref.\cite{Calabrese:2001kj,Frittelli:1996nj}.

As far as we know, there are many papers studying the constraint preserving boundary condition formulation. Based on its hyperbolic structure,  a set of constraint preserving boundary conditions are introduced  for the BSSN formulation of the Einstein evolution equations in spherical symmetry~\cite{Alcubierre:2014joa}. The case of the first-order version of the Z4 system is considered, and constraint-preserving boundary conditions of the Sommerfeld type are provided in~\cite{Bona:2004ky}. The Einstein–Christoffel formulation of the Einstein equations in the case of
spherical symmetry are discussed in~\cite{Calabrese:2001kj,Frittelli:2003ym}. Constraint preserving boundary condition formulation in the Eddington-Finkelstein coordinates can be found in~\cite{Iriondo:2001ig}. 
Recently, the sufficient conditions required for the PDEs of the system to guarantee the constraint preservation are studied in~\cite{Abalos:2021rqs}, more meticulously.

Under the restriction of spherical symmetry, a constraint preserving formulation of the ADM problem for the Einstein equations  transformed from the Bondi-Sachs formulation has been provided in the Ref.\cite{Frittelli:2006ap,Frittelli:2007jb}. 
After adding a scalar field, the Einstein equations coupled to the scalar field in spherical symmetry are cast into a symmetric-hyperbolic system of equations in which the scalar field, lapse, and shift are fundamental variables. Without the spherical symmetry,
the hyperbolicity of the Bondi-like gauge is studied in the Ref.\cite{Giannakopoulos:2020dih,Giannakopoulos:2021pnh}.

In our paper, we adhere to the assumption of spherical symmetry, and extend the above Bondi-like formulation to the case with a cosmological constant so as to study the instability of AdS spacetime. 
We use the constraint preserving formulation to give a numerical evidence that the pure AdS spacetime is unstable under slight perturbations. It should be emphasized that our slight perturbations are located at the outer boundary in our formulation 
which is different from the one where the perturbation happens in the initial data. The relationship between the perturbation and the position of the apparent horizon is given numerically.
This work reveals the instability of AdS spacetime from another perspective.

The paper is organized as follows. In Sec.\ref{sec_2}, we present the main evolution equations and propagation of the constraints by using the Bianchi identity. Numerical simulations is displayed in Sec.\ref{sec_3}. 
We consider two different small perturbations and get the similar conclusions. 
We perform the numerical convergence test in Sec.\ref{sec_4} and show that our numerical scheme is second order self convergent as we use the second order Runge-Kutta in time integrator and the second order special difference algorithm.
Sec.\ref{sec_5} is the conclusion and the discussion.

\section{main evolution equations and propagation of the constraints}\label{sec_2}
The system  we are considering is the same as the one in~\cite{Bizon:2011gg}. The action is given by
\begin{eqnarray}\label{action}
	S=\frac{1}{16\pi G}\int\mathrm{d}^4x\sqrt{-g}\Big(R-2\Lambda-\nabla_a\varphi\nabla^a\varphi\Big)\, ,
\end{eqnarray}
where $R$ is the Ricci scalar, $g$ is the determinant of the metric $g_{ab}$, $G$ is the gravitational constant, $\varphi$ is a scalar field, $\Lambda$ is the cosmological constant. Varying the action (\ref{action}), we get the equations of motion which can be expressed as
\begin{eqnarray}\label{EOM_g}
	R_{ab}-\frac12g_{ab}R+\Lambda g_{ab} = 8\pi T_{ab}\, ,
\end{eqnarray}
and
\begin{eqnarray}\label{EOM_varphi}
	\nabla_a\nabla^a\varphi=0\, ,
\end{eqnarray}
where $T_{ab}$ is the stress-energy tensor, which has a form
\begin{eqnarray}\label{stress-tensor}
	T_{ab}=\nabla_a\varphi\nabla_b\varphi-\frac12g_{ab}\nabla_c\varphi\nabla^c\varphi\, .
\end{eqnarray}
We report here that the spherically symmetric ADM formulation is motivated by the Bondi-Sachs formulation. The line element in the coordinate system $(v,\hat{r},\theta,\phi)$ adapted to the null slicing takes the form
\begin{eqnarray}\label{metric_Bondi}
	\mathrm{d}s^2=-e^{2\beta}\frac{V}{\hat{r}}\mathrm{d}v^2+2e^{2\beta}\mathrm{d}v\mathrm{d}\hat{r}+\hat{r}^2(\mathrm{d}\theta^2+\sin^2\theta\mathrm{d}\phi^2)\, ,
\end{eqnarray}
where $\beta=\beta(v,\hat{r})$ and $V=V(v,\hat{r})$ are two undetermined functions. For convenience, the field equations are denoted in a more compact form
\begin{eqnarray}
	E_{ab}\equiv G_{ab}+\Lambda g_{ab}-8\pi T_{ab}=0\, .
\end{eqnarray}
Based on the Bianchi identity, the independent components out of $E_{ab}$ are discussed in Appendix \ref{Bianchi identity}. Therefore, one gets the  equations of evolution
\begin{eqnarray}
	E_{\hat{r}\hat{r}}=0\, ,\quad E_{v\hat{r}}=0\, ,\quad\nabla_a\nabla^a\varphi=0\, ,
\end{eqnarray}
and one constraint
\begin{eqnarray}
	E_v{}^{\hat{r}}=0\, .
\end{eqnarray}
Substituting the metric (\ref{metric_Bondi}) into the above equations, we get
\begin{eqnarray}
	&&\beta_{,\hat{r}}-2\pi\hat{r}(\varphi_{,\hat{r}})^2=0\, ,\nonumber\\
	&&V_{,\hat{r}}-e^{2\beta}(1-\Lambda\hat{r}^2)=0\, ,\nonumber \\ 
	&&2\varphi_{,v\hat{r}}+\frac{V}{\hat{r}}\varphi_{,\hat{r}\hat{r}}+\Big(\frac{V_{,\hat{r}}}{\hat{r}}+\frac{V}{\hat{r}^2}\Big)\varphi_{,\hat{r}}+\frac{2}{\hat{r}}\varphi_{,v}=0\, ,
\end{eqnarray}
where the second equation comes from $$g^{\hat{r}\hat{r}}E_{\hat{r}\hat{r}}+2g^{v\hat{r}}E_{v\hat{r}}=0.$$ The constraint variable $E_v{}^{\hat{r}}$ has the form
\begin{eqnarray}
	E_{v}{}^{\hat{r}}\equiv G_{v}{}^{\hat{r}}+\Lambda\delta_{v}{}^{\hat{r}}-8\pi T_{v}{}^{\hat{r}}=\frac{e^{-2\beta}}{\hat{r}^2}\Big[2V\beta_{,v}-V_{,v}-8\pi\hat{r}^2\Big(\varphi_{,v}\varphi_{,v}+\frac{V}{\hat{r}}\varphi_{,v}\varphi_{,\hat{r}}\Big)\Big]\, .
\end{eqnarray}
The essential difference between the Bondi-Sachs problem  and  the ADM problem is that one of them uses null slices, which is unique in spherical symmetry, whereas the other one uses spacelike slices, which is not unique. At every point of the spacetime there is one ingoing or outgoing null tangent vector, but there is a one-parameter family of spacelike vectors. The parameter corresponds to the ``slope" of the slice at that point.  Without losing generality, the relevant ADM formulation is obtained by making a coordinate transformation~\cite{Frittelli:2006ap,Frittelli:2007jb}
\begin{eqnarray}
	\hat{r}&=&r\, ,\nonumber\\
	v&=&t+f(r)\, .\label{coordinate_transformation}
\end{eqnarray}
Clearly, one gets a unique relationship between the Bondi-Sachs and ADM variables, if one fixes the function $f(r)$. Under the coordinate transformation (\ref{coordinate_transformation}), the Bondi-Sachs metric (\ref{metric_Bondi}) has the form
\begin{eqnarray}
	\mathrm{d}s^2=-\frac{e^{2\beta}}{f^{\prime}(2-f^{\prime}V/r)}\mathrm{d}t^2+e^{2\beta}f^{\prime}(2-f^{\prime}V/r)\Big[\mathrm{d}r+\frac{(1-f^{\prime}V/r)}{f^{\prime}(2-f^{\prime}V/r)}\mathrm{d}t\Big]^2+r^2(\mathrm{d}\theta^2+\sin^2\theta\mathrm{d}\phi^2)\, .
\end{eqnarray}
Comparing with the standard ADM metric,
\begin{eqnarray}
	\mathrm{d}s^2=-\alpha^2\mathrm{d}t^2+\gamma_{rr}(\mathrm{d}r+\beta^r\mathrm{d}t)^2+\gamma_Tr^2(\mathrm{d}\theta^2+\sin^2\theta\mathrm{d}\phi^2)\, ,
\end{eqnarray}
one has the identifications
\begin{eqnarray}
	\alpha^2&=&\frac{e^{2\beta}}{f^{\prime}(2-f^{\prime}V/r)}\, ,\nonumber\\
	\beta^r&=&\frac{1-f^{\prime}V/r}{f^{\prime}(2-f^{\prime}V/r)}\, ,\label{Bondi_identifications}
\end{eqnarray}
with
\begin{eqnarray}
	\gamma_T&=&1\, ,\nonumber\\
	\gamma_{rr}&=&(f^{\prime})^2\frac{\alpha^2}{(1-f^{\prime}\beta^r)^2}\, ,\label{advanced_Bondi_Sachs_gauge}
\end{eqnarray}
where $\alpha$ and $\beta^r$ are referred to as the lapse and shift, $f^{\prime}\equiv \mathrm{d}f(r)/\mathrm{d}r$. Eqs.(\ref{advanced_Bondi_Sachs_gauge}) are called as the advanced Bondi-Sachs gauge. In the followings, we will set $f(r)=r$. This slicing corresponds to what is usually referred to as Kerr-Schild coordinates~\cite{Kidder:2000yq}.

In order to obtain an initial value problem in a time slicing, we transform the coordinates from $(v,\hat{r})$ into $(t,r)$ by using Eq.(\ref{coordinate_transformation}) and the variables from $(\beta,V)$ to $(\alpha,\beta^r)$ with the help of Eq.(\ref{Bondi_identifications}). After some calculation, the initial value problem is found
\begin{eqnarray}
	\dot{\alpha}&=&\alpha_{,r}+\frac{\alpha(1-2\beta^r)}{2r}-\frac{\alpha^3}{2r}(1-\Lambda r^2)-2\pi r\alpha(P-Q)^2\, ,\nonumber\\
	\dot{\beta}^r&=&\beta^r_{,r}-\frac{(1-\beta^r)(1-2\beta^r)}{r}+\frac{1-\beta^r}{r}\alpha^2(1-\Lambda r^2)\, ,\nonumber\\
	\dot{P}&=&2\beta^rP_{,r}+(1-2\beta^r)Q_{,r}+\frac{2(1-\beta^r)}{r}P+\Big(\frac{1-\Lambda r^2}{r}\alpha^2+\frac{1-2\beta^r}{r}\Big)(Q-P)\, ,\label{evolution}\nonumber\\
	\dot{Q}&=&P_{,r}\, ,\nonumber\\
	\dot{\varphi}&=&P\, ,
\end{eqnarray}
where, to reduce the order of the equation of the scalar field $\varphi$, we define the derivatives of the field $\varphi$ as new variables
\begin{eqnarray}
	P&\equiv&\dot{\varphi}\, ,\nonumber\\
	Q&\equiv&\varphi_{,r}\, .\label{Q}
\end{eqnarray}
Since our numerical simulation is performed on the region $[r_{\text{min}},r_{\text{max}}]$, in order to construct the initial boundary value problem, the boundary conditions have to be considered. To all intents and purposes, the constraint is used to give appropriate boundary conditions, and can be written as
\begin{eqnarray}
	E_v{}^{\hat{r}}\equiv\mathcal{C}_1\equiv\frac{2(1-2\beta^r)}{r\alpha^3}\alpha_{,r}+\frac{2}{r\alpha^2}\beta^r_{,r}+\frac{\alpha^2-1+2\beta^r}{r^2\alpha^2}-\Lambda-\frac{4\pi}{\alpha^2}\Big[P^2+(1-2\beta^r)Q^2\Big]=0\, .
\end{eqnarray}
This is a first-order differential constraint on the lapse and shift. From Appendix \ref{Bianchi identity}[see Eq.(\ref{E_v^r_3})], constraint variable $\mathcal{C}_1$ satisfies the following equation
\begin{eqnarray}
	\dot{\mathcal{C}}_1-\mathcal{C}_{1,r}+\cdots=0\, ,
\end{eqnarray}
where $``\cdots"$ represent undifferentiated terms. Therefore, the constraint variable $\mathcal{C}_1$ propagates at the characteristic speed $v=1$. Because our numerical evolution is performed in a finite spatial region $[r_{\text{min}},r_{\text{max}}]$, 
the ingoing constraint $\mathcal{C}_1=0$ must be imposed on the outer boundary $r_{\text{max}}$~\cite{Calabrese:2001kj}. Note that the definition of $Q$, i.e., Eq.(\ref{Q}), provides another constraint which is denoted as $\mathcal{C}_2=0$. But the characteristic speed of this constraint is zero, so the boundary conditions is not necessary.  

To illustrate the requirement of the boundary conditions for the numerical simulations, diagonalization of principal part is carried out. First, the system of equations (\ref{evolution}) has a block-diagonal principal part which can be written as a matrix form
\begin{eqnarray}
	M=\begin{bmatrix}
		1 & 0 & 0        & 0          & 0 \\
		0 & 1 & 0        & 0          & 0 \\
		0 & 0 & 2\beta^r & 1-2\beta^r & 0 \\
		0 & 0 & 1        & 0          & 0 \\
		0 & 0 & 0        & 0          & 0
	\end{bmatrix}\, .\label{pricinpal_part}
\end{eqnarray}
The eigenvalues of this matrix are $$1,\quad 1\, ,\quad 1,\quad -(1-2\beta^r),\quad 0.$$ It is not hard to find that the characteristic fields are 
$$ \alpha, \quad  \beta^r, \quad U^{-}, \quad  U^{+}, \quad  \varphi, $$
where
\begin{eqnarray}
U^-&=& P+(1-2\beta^r)Q\, ,\nonumber\\
U^{+}&=&P-Q\, .
\end{eqnarray}
The speed of $\alpha$, $\beta^r$, $U^{-}$ is $1$, the speed of $U^{+}$ is $2\beta^r-1$ and the speed of $\varphi$ is $0$. Due to the fact that the characteristic fields span the whole eigenspace. It means that the system (\ref{evolution}) is strongly hyperbolic~\cite{Sarbach:2012pr}. Hence, after diagonalization, in terms of these characteristic fields, the system (\ref{evolution}) reads
\begin{eqnarray}
\label{evolution_characteristic}
	\dot{\alpha}&=&\alpha_{,r}+\frac{\alpha(1-2\beta^r)}{2r}-\frac{\alpha^3}{2r}(1-\Lambda r^2)-2\pi r\alpha(U^{+})^2\, ,\nonumber\\
	\dot{\beta}^r&=&\beta^r_{,r}-\frac{(1-\beta^r)(1-2\beta^r)}{r}+\frac{1-\beta^r}{r}\alpha^2(1-\Lambda r^2)\, ,\nonumber\\
	\dot{U}^{+}&=&-(1-2\beta^r)U^{+}_{,r}+\frac{U^{-}}{r}-\frac{1-\Lambda r^2}{r}\alpha^2U^{+}\, ,\nonumber\\
	\dot{U}^{-}&=&U^{-}_{,r}+\frac{U^{-}}{r}-\frac{1-\Lambda r^2}{r}\alpha^2U^{+}+(U^{+}-U^{-})\frac{\alpha^2(1-\Lambda r^2)-1+2\beta^r}{r}\, ,\nonumber\\
	\dot{\varphi}&=&\frac{(1-2\beta^r)U^{+}+U^{-}}{2(1-\beta^r)}\, .
	\end{eqnarray}
Two initial constraints $(\mathcal{C}_1=0, \mathcal{C}_2=0)$ in terms of characteristic fields become
\begin{eqnarray}
	\frac{2(1-2\beta^r)}{r\alpha^3}\alpha_{,r}+\frac{2}{r\alpha^2}\beta^r_{,r}+\frac{\alpha^2-1+2\beta^r}{r^2\alpha^2}-\Lambda-\frac{2\pi}{\alpha^2}\Big[\frac{(U^{-})^2+(1-2\beta^r)(U^{+})^2}{1-\beta^r}\Big]=0\, ,\label{constraint_1}
\end{eqnarray}
and
\begin{eqnarray}
	\varphi_{,r}-\frac{-U^{+}+U^{-}}{2(1-\beta^r)}=0\, ,\label{constraint_2}
\end{eqnarray}
respectively. On account of the positive characteristic speed of the constraint variable $\mathcal{C}_1$, $\mathcal{C}_1=0$ needs to be enforced at the outer boundary $r_{\text{max}}$. However, since our numerical simulation is a free evolution, $\mathcal{C}_1=0$ should be transformed in order to achieve the convenience of the numerical discretization. Trading spatial derivatives for time derivatives, Eq.(\ref{constraint_1}) in turn becomes
\begin{eqnarray}
	\dot{\beta}^r+\frac{1-2\beta^r}{\alpha}\dot{\alpha}-\pi r\frac{(1-2\beta^r)U^{+}+U^{-}}{1-\beta^r}\Big[(2\beta^r-1)U^{+}+U^{-}\Big]=0\, .\label{boundary_condition}
\end{eqnarray}
Since the characteristic fields $\alpha$, $\beta^r$, $U^{-}$ get into the computing domain at the outer boundary, so the values of these fields are specified freely. However, they are restricted to
\begin{eqnarray}
	\mathcal{C}_1|_{r_{\text{max}}}=0\, .
\end{eqnarray}
Actually, there are only two freely chosen variables among $\alpha$, $\beta^r$, $U^{-}$. In order to perform the scalar field perturbation entering the computational domain $[r_{\text{min}},r_{\text{max}}]$, we enforce the constraint-preserving boundary condition by solving for $\alpha$, given $\beta^r$, $U^{+}$, $U^{-}$ at the outer boundary. From the boundary condition (\ref{boundary_condition}), we have
\begin{eqnarray}
	\dot{\alpha}=-\frac{\alpha}{1-2\beta^r}\dot{\beta}^r+\pi r\alpha\frac{(U^{-})^2-(1-2\beta^r)^2(U^{+})^2}{(1-2\beta^r)(1-\beta^r)}\, .
\end{eqnarray}
For the inner boundary, we know that the characteristic variable $U^{+}$ has characteristic speed $2\beta^r-1$. When $\beta^r<1/2$, or $2\beta^r-1<0$, strictly speaking, we should add an inner boundary condition for the characteristic variable $U^{+}$. But in our simulation, extrapolation is always used at the inner boundary whenever $U^{+}$ is an ingoing mode or an outgoing mode. For actual simulations, some reasonable assumption is supposed to be made, whose validity can only be verified after the fact~\cite{Frittelli:2006ap,Frittelli:2007jb}.

Our simulation tracks whether the apparent horizon appears. In Bondi-Sachs gauge, the expansion $\Theta$ is~\cite{2010Numerical}
\begin{eqnarray}
	\Theta=\frac{\sqrt{2}}{r\sqrt{\gamma_T}}\Big[\frac{1}{\alpha}\partial_t(\sqrt{\gamma_T}r)+\Big(\frac{1}{\sqrt{\gamma_{rr}}}-\frac{\beta^r}{\alpha}\Big)\partial_r(\sqrt{\gamma_T}r)\Big]=\frac{\sqrt{2}(1-2\beta^r)}{r\alpha}\, .
\end{eqnarray}
Therefore, the location of the apparent horizon is given by $\beta^r(r_H)=1/2$. Numerical simulations will be shown in the next section.  Until now, we have constructed the initial boundary value problem which is suitable for numerical calculation.

\section{numerical simulations}\label{sec_3}
To implement the proposed boundary treatment, a straightforward second-order dissipative method of lines (MOL) is chosen. Spatial derivatives are discretized with second-order centered differences plus fourth-order dissipation as discussed in~\cite{2006Introduction}, while for the time integrator we use second order Runge-Kutta. Our uniform grid structure consists of points $i=1,\cdots,N$, with grid spacing $\Delta r=L/(N-1)$, where $L=r_{\text{max}}-r_{\text{min}}$. Spatial derivatives are applied according to standard formulas~\cite{Calabrese:2001kj}
\begin{eqnarray}
	Mf^{\prime}&\to& MD_0f-\frac{\epsilon_i}{\Delta t}(\Delta r)^4D_{+}D_{+}D_{-}D_{-}f\, ,
\end{eqnarray}
where
\begin{eqnarray}	
	(D_0f)_i&=&\frac{f_{i+1}-f_{i-1}}{2\Delta r}\, ,\nonumber\\
	(D_+f)_i&=&\frac{f_{i+1}-f_{i}}{\Delta r}\, ,\nonumber\\
	(D_-f)_i&=&\frac{f_{i}-f_{i-1}}{\Delta r}\, ,
\end{eqnarray}
and
\begin{eqnarray}
	-\frac{\epsilon_i}{\Delta t}(\Delta r)^4D_{+}D_{+}D_{-}D_{-}f=-\frac{\epsilon_i}{\Delta t}(f_{i+2}-4f_{i+1}+6f_{i}-4f_{i-1}+f_{i-2})\, ,
\end{eqnarray}
$M$ is given by Eq.(\ref{pricinpal_part}),  and $\epsilon_i$ is the dissipative factor. In order to evaluate derivatives at the boundaries, ghost zones which are artificial points beyond the boundaries where field values are defined via extrapolation (second order one-sided difference is actually the same as second order central difference with a third order extrapolation). Spatial indicators  of these ghost points are $i=0$ and $i=N+1$. Field values at these ghost points are defined via third order extrapolations and the same derivative operator is applied at $i=1,\cdots,N$. After using standard finite differences for the spatial derivatives, we can rewrite our original differential equations (\ref{evolution_characteristic}) as a coupled system of ordinary differential equations of the form
\begin{eqnarray}
	\frac{\mathrm{d}u}{\mathrm{d}t}=S(u,t)\, ,\label{odes}
\end{eqnarray}
where $u$ is a vector constructed from the values of the function $u$ in the spatial grid points. The second order Runge-Kutta algorithm used in our simulation takes the form
\begin{eqnarray}
	u^{*}&=&u^{n}+\Delta tS(u^n,t^n)/2\, ,\nonumber\\
	u^{n+1}&=&u^n+\Delta tS(u^{*},t^n+\Delta t/2)\, ,
\end{eqnarray}
where $u^n=u(t^n)$ and $t^n=n\Delta t$.

For our simulation, $u$ is a $5N-2$ dimensional vector. They are $$\alpha_1\, ,\cdots\, ,\alpha_N\, , \beta^r_1\, ,\cdots\, ,\beta^r_{N-1}\, , U^{+}_1\, ,\cdots\, ,U^{+}_N\, , U^{-}_1\, ,\cdots\, ,U^{-}_{N-1}\, , \varphi_1\, ,\cdots\, ,\varphi_N\, ,$$ 
which are variables of MOL. 
$\beta^r_N$ and $U^{-}_{N}$ are the free variables at the outer boundary. They do not iterate in the RK algorithm. Since the time-integration method is explicit, the time-step is limited by the Courant-Friedrichs-Lewy (CFL) condition. The CFL condition takes the form as follow
\begin{eqnarray}
	\text{max}|\lambda_a|\Delta t\le\Delta r\, ,
\end{eqnarray}
where $\lambda_a$ is the characteristic speed.  We  choose $\Delta t/\Delta r=0.25$, it  is stable enough for the scheme. It should be pointed out that since four order derivative cannot be obtained at the boundary points $i=1$, $i=N$, so there is no dissipation at these points. It means that $\epsilon_1=\epsilon_N=0$. For the dissipation factor $\epsilon_i$, it is chosen based on a von Neumann stability analysis~\cite{2006Introduction}. We choose $\epsilon_i=0.6/16$.
Now, we write the form of ordinary differential equations (\ref{odes}) definitely. For $i=2,\cdots,N-1$,
\begin{eqnarray}
	\dot{\alpha}_i&=&\frac{\alpha_{i+1}-\alpha_{i-1}}{2\Delta r}+\Big[\frac{\alpha(1-2\beta^r)}{2r}-\frac{\alpha^3}{2r}(1-\Lambda r^2)-2\pi r\alpha(U^{+})^2\Big]_i\nonumber\\
	&&-\frac{\epsilon_i}{\Delta t}(\alpha_{i+2}-4\alpha_{i+1}+6\alpha_{i}-4\alpha_{i-1}+\alpha_{i-2})\, ,\nonumber\\
	\dot{\beta}^r_i&=&\frac{\beta^r_{i+1}-\beta^r_{i-1}}{2\Delta r}+\Big[-\frac{(1-\beta^r)(1-2\beta^r)}{r}+\frac{1-\beta^r}{r}\alpha^2(1-\Lambda r^2)\Big]_i\nonumber\\
	&&-\frac{\epsilon_i}{\Delta t}(\beta^r_{i+2}-4\beta^r_{i+1}+6\beta^r_{i}-4\beta^r_{i-1}+\beta^r_{i-2})\, ,\nonumber\\
	\dot{U}^{+}_i&=&-(1-2\beta^r_i)\frac{U^{+}_{i+1}-U^{+}_{i-1}}{2\Delta r}+\Big(\frac{U^{-}}{r}-\frac{1-\Lambda r^2}{r}\alpha^2U^{+}\Big)_i\nonumber\\
	&&-\frac{\epsilon_i}{\Delta t}(U^{+}_{i+2}-4U^{+}_{i+1}+6U^{+}_{i}-4U^{+}_{i-1}+U^{+}_{i-2})\, ,\nonumber\\
	\dot{U}^{-}_i&=&\frac{U^{-}_{i+1}-U^{-}_{i-1}}{2\Delta r}+\Big[\frac{U^{-}}{r}-\frac{1-\Lambda r^2}{r}\alpha^2U^{+}+(U^{+}-U^{-})\frac{\alpha^2(1-\Lambda r^2)-1+2\beta^r}{r}\Big]_i\nonumber\\
	&&-\frac{\epsilon_i}{\Delta t}(U^{-}_{i+2}-4U^{-}_{i+1}+6U^{-}_{i}-4U^{-}_{i-1}+U^{-}_{i-2})\, ,\nonumber\\
	\dot{\varphi}_i&=&\Big[\frac{(1-2\beta^r)U^{+}+U^{-}}{2(1-\beta^r)}\Big]_i-\frac{\epsilon_i}{\Delta t}(\varphi_{i+2}-4\varphi_{i+1}+6\varphi_{i}-4\varphi_{i-1}+\varphi_{i-2})\, .
\end{eqnarray}
We have pointed out that for the inner boundary, the values of ghost points are obtained by interpolation. Therefore, at the inner boundary $i=1$,
\begin{eqnarray}
	\dot{\alpha}_1&=&\frac{\alpha_{2}-\alpha_{0}}{2\Delta r}+\Big[\frac{\alpha(1-2\beta^r)}{2r}-\frac{\alpha^3}{2r}(1-\Lambda r^2)-2\pi r\alpha(U^{+})^2\Big]_1\, ,\nonumber\\
	\dot{\beta}^r_1&=&\frac{\beta^r_{2}-\beta^r_{0}}{2\Delta r}+\Big[-\frac{(1-\beta^r)(1-2\beta^r)}{r}+\frac{1-\beta^r}{r}\alpha^2(1-\Lambda r^2)\Big]_1\, ,\nonumber\\
	\dot{U}^{+}_1&=&-(1-2\beta^r_1)\frac{U^{+}_{2}-U^{+}_{0}}{2\Delta r}+\Big(\frac{U^{-}}{r}-\frac{1-\Lambda r^2}{r}\alpha^2U^{+}\Big)_1\, ,\nonumber\\
	\dot{U}^{-}_1&=&\frac{U^{-}_{2}-U^{-}_{0}}{2\Delta r}+\Big[\frac{U^{-}}{r}-\frac{1-\Lambda r^2}{r}\alpha^2U^{+}+(U^{+}-U^{-})\frac{\alpha^2(1-\Lambda r^2)-1+2\beta^r}{r}\Big]_1\, ,\nonumber\\
	\dot{\varphi}_1&=&\Big[\frac{(1-2\beta^r)U^{+}+U^{-}}{2(1-\beta^r)}\Big]_1\, ,
\end{eqnarray}
where
\begin{eqnarray}
	\alpha_0&=&3\alpha_1-3\alpha_2+\alpha_3\, ,\nonumber\\
	\beta^r_0&=&3\beta^r_1-3\beta^r_2+\beta^r_3\, ,\nonumber\\
	U^{+}_0&=&3U^{+}_1-3U^{+}_2+U^{+}_3\, ,\nonumber\\
	U^{-}_0&=&3U^{-}_1-3U^{-}_2+U^{-}_3\, ,\nonumber\\
	\varphi_0&=&3\varphi_1-3\varphi_2+\varphi_3\, .
\end{eqnarray}
At the outer boundary, since only the characteristic field $U^{+}$ leaves out of the computing domain, we extrapolate $U^{+}$ at the ghost zone point $i=N+1$. That is
\begin{eqnarray}
	U^{+}_{N+1}=U^{+}_{N-2}-3U^{+}_{N-1}+3U^{+}_N\, .
\end{eqnarray}
So at $i=N$,
\begin{eqnarray}
	\dot{U}^{+}_N&=&-(1-2\beta^r_N)\frac{U^{+}_{N+1}-U^{+}_{N-1}}{2\Delta r}+\Big(\frac{U^{-}}{r}-\frac{1-\Lambda r^2}{r}\alpha^2U^{+}\Big)_N\, ,
\end{eqnarray}
and
\begin{eqnarray}
	\dot{\varphi}_N=\Big[\frac{(1-2\beta^r)U^{+}+U^{-}}{2(1-\beta^r)}\Big]_N\, .
\end{eqnarray}
In fact, there are only two freely chosen variables among $\alpha$, $\beta^r$ and $U^{-}$. From the boundary condition, we have
\begin{eqnarray}
	\dot{\alpha}_N=-\Big[\frac{\alpha}{1-2\beta^r}\Big]_N\dot{\beta}^r_N+\Big[\pi r\alpha\frac{(U^{-})^2-(1-2\beta^r)^2(U^{+})^2}{(1-2\beta^r)(1-\beta^r)}\Big]_N\, .
\end{eqnarray}
This is nothing but the evolution of $\alpha_N$. Our simulation is to provide numerical evidence for the instability of the pure AdS spacetime.  
We consider a small perturbation of the scalar field  $U^{-}$ on the outer boundary. The reason is that $U^-$ is the only ingoing mode on the outer boundary.

Before choosing the boundary condition for $\beta^r_N$ and $U^{-}_N$, it is worth mentioning that in our coordinate (Bondi-Sachs gauge), the pure AdS solution can be expressed as
\begin{eqnarray}
	\alpha(r)=\frac{1}{\sqrt{1-\frac{2\Lambda}{3}r_{\text{max}}^2+\frac{\Lambda}{3}r^2}}\, ,\quad \beta^r(r)=\frac{2\Lambda(r^2-r_{\text{max}}^2)}{6+2\Lambda r^2-4\Lambda r_{\text{max}}^2}\, .\label{pure_AdS}
\end{eqnarray}
Actually, under the above Eq.(\ref{pure_AdS}) of $\alpha$ and $\beta^r$, the metric becomes
\begin{eqnarray}
	\mathrm{d}s^2=\frac{3(\Lambda r^2-3)}{(\Lambda r_{\text{max}}^2-3)^2}\mathrm{d}v^2+\frac{6}{3-\Lambda r_{\text{max}}^2}\mathrm{d}v\mathrm{d}r+r^2(\mathrm{d}\theta^2+\sin^2\theta\mathrm{d}\phi^2)\, .
\end{eqnarray}
Note that $3/(3-\Lambda r_{\text{max}}^2)>0$ is a constant. Define 
$$\mathrm{d}t^{\prime}+\frac{3 \mathrm{d}r}{(3-\Lambda r^2)}=\frac{3\mathrm{d}v}{(3-\Lambda r_{\text{max}}^2)}\, ,$$
 then the metric becomes
\begin{eqnarray}
	\mathrm{d}s^2=-\Big(1-\frac{\Lambda r^2}{3}\Big)\mathrm{d}t^{\prime 2}+\Big(1-\frac{\Lambda r^2}{3}\Big)^{-1}\mathrm{d}r^2+r^2(\mathrm{d}\theta^2+\sin^2\theta\mathrm{d}\phi^2)\, .
\end{eqnarray}
This is the pure AdS spacetime solution in static coordinates. It can be shown that $\beta^r<1/2$, as $r\in[0,r_{\text{max}}]$. Therefore, this pure AdS solution has no apparent horizon as we all know. 

\begin{figure}[htbp]
	\centering
	\includegraphics[width=3.0 in]{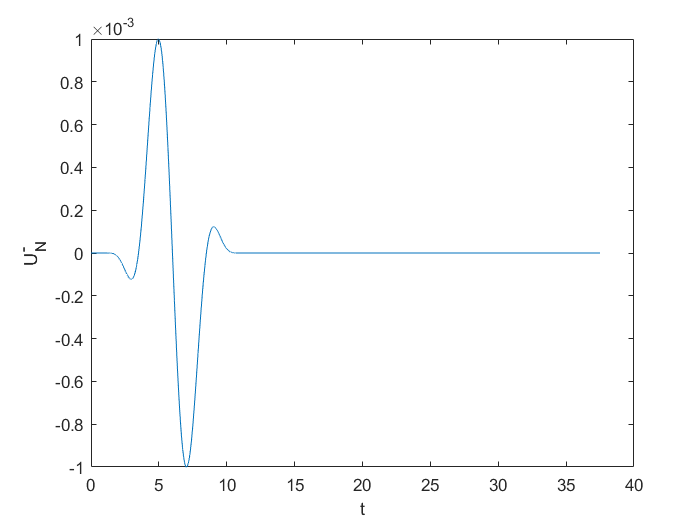}
	\caption{The configuration of $U^{-}_N(t)$ with $t_\text{I}=1$, $t_\text{F}=11$, $A=1\times10^{-3}$ for the case of type \uppercase\expandafter{\romannumeral1}.}
	\label{fig_Un}
\end{figure}

\begin{figure}[htbp]
	\centering
	\includegraphics[width=3.0 in]{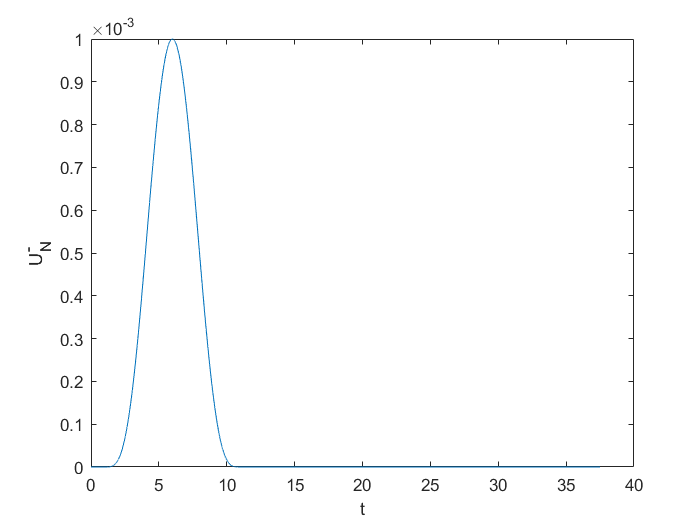}
	\caption{The configuration of $U^{-}_N(t)$ with $t_\text{I}=1$, $t_\text{F}=11$, $A=1\times10^{-3}$ for the case of type \uppercase\expandafter{\romannumeral2}.}
	\label{fig_Un_2}
\end{figure}

Now, we start to consider the boundary condition for $\beta^r_N$ and $U^{-}_N$. We choose
\begin{eqnarray}
	\beta^r_N(t)=\beta^r_N(0)\, .
\end{eqnarray}
This is a Dirichlet boundary condition for $\beta^r_N(t)$, since $\beta^r$ is a characteristic field at outer boundary with a speed 1. But for $U^{-}_{N}$, it is described by
\begin{eqnarray}
	U^{-}_N(t)=\left\{
	\begin{aligned}
	&A\cdot N_1\Big(\frac{t-t_\text{I}}{t_\text{F}-t_\text{I}}\Big)^4\Big(1-\frac{t-t_\text{I}}{t_\text{F}-t_\text{I}}\Big)^4\sin\Big[\pi\Big(4\frac{t-t_{\text{I}}}{t_{\text{F}}-t_{\text{I}}}+1\Big)\Big]&\, ,\quad\text{type \uppercase\expandafter{\romannumeral1}} \\
	&A\cdot N_2\Big(\frac{t-t_\text{I}}{t_\text{F}-t_\text{I}}\Big)^4\Big(1-\frac{t-t_\text{I}}{t_\text{F}-t_\text{I}}\Big)^4&\, ,\quad\text{type \uppercase\expandafter{\romannumeral2}}
	\end{aligned}
	\right.
\end{eqnarray}
if $t\in[t_\text{I},t_\text{F}]$, and $U^{-}_N(t)=0$ otherwise. $N_1$, $N_2$ are the normalization factors and their values are $316.494$ and $256$. We illustrate $U^{-}_N(t)$ at type \uppercase\expandafter{\romannumeral1} in Fig.\ref{fig_Un} and $U^{-}_N(t)$ at type \uppercase\expandafter{\romannumeral2} in Fig.\ref{fig_Un_2}. Therefore, we construct initial data by setting
\begin{eqnarray}
	\alpha(r)=\frac{1}{\sqrt{1-\frac{2\Lambda}{3}r_{\text{max}}^2+\frac{\Lambda}{3}r^2}}\, ,\quad \beta^r(r)=\frac{2\Lambda(r^2-r_{\text{max}}^2)}{6+2\Lambda r^2-4\Lambda r_{\text{max}}^2}\, ,\quad U^{+}=U^{-}=0\, ,\quad \varphi=1\, .\label{initial_data}
\end{eqnarray}
This means that the initial metric configuration is the pure AdS metric. In our numerical simulation, we always choose the pure AdS solution as the initial data but adjust the form of the ingoing characteristic field $U^{-}$ at outer boundary.

\begin{figure}[htbp]
	\centering
	\includegraphics[width=3.0 in]{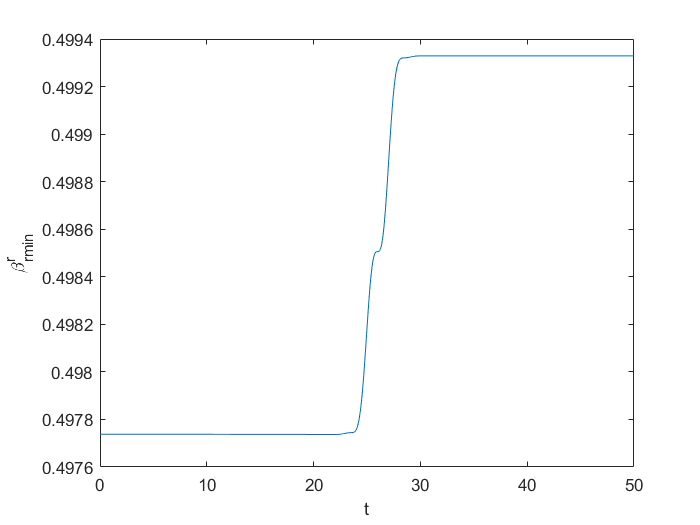}
	\caption{In this case of type \uppercase\expandafter{\romannumeral1}, the value of the shift function $\beta^r$ changes with time at $r_{\text{min}}=1$ with outer boundary $r_\text{max}=21$. There is no apparent horizon in $[r_{\text{min}}, r_{\text{max}}]$.}
	\label{fig1}
\end{figure}

\begin{figure}[htbp]
	\centering
	\includegraphics[width=3.0 in]{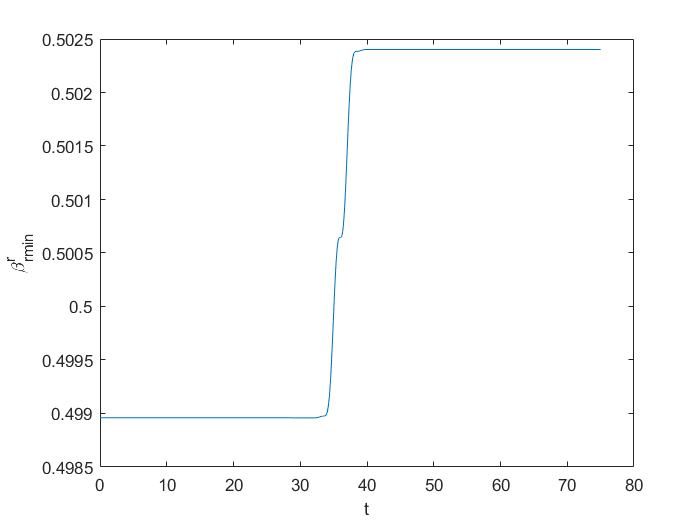}
	\caption{In this case of type \uppercase\expandafter{\romannumeral1}, the value of the shift function $\beta^r$ changes with time at $r_{\text{min}}=1$ with outer boundary $r_\text{max}=31$. There is a apparent horizon in $[r_{\text{min}}, r_{\text{max}}]$.}
	\label{fig2}
\end{figure}

Here, in our simulation, we choose $\Lambda=-1$, $r_{\text{min}}=1$, $t_\text{I}=1$, $t_\text{F}=11$. In the case of type \uppercase\expandafter{\romannumeral1}, the amplitude of $U^{-}_N$ is chosen by $A=1\times10^{-3}$. 
The number of spatial points we choose is $N=4001$. Increasing the position of disturbance (from $r_\text{max}=21$ to $r_\text{max}=31$), we track the position of the apparent horizon. 
For the final distribution of $\beta^r$,  both results show that $\beta^r$ decreases as  $r$ increases. 
Moreover, the value of $\beta^r$ for the final moment reaches the maximum at $r_\text{min}$ and gradually decreases to $0$ (this is required by the Dirichlet boundary condition we adopted.). 
These results are shown on the following two figures, see Fig.\ref{fig1} and Fig.\ref{fig2}. In Fig.\ref{fig1}, we find that there is no apparent horizon in $[r_{\text{min}}, r_{\text{max}}]$. In Fig.\ref{fig2}, we find that there is an apparent horizon in $[r_{\text{min}}, r_{\text{max}}]$.

It should be pointed out that even if there is no apparent horizon in $[r_{\text{min}}, r_{\text{max}}]$, we can not completely rule out the situation that there is a apparent horizon in $[0,r_{\text{min}}]$. 
However, this does not affect our conclusion. Actually, when one sets $r_\text{min}=0.5$, $r_\text{max}=21$, one can find the apparent horizon is located at $r_H=0.7768<1$ with the same $U_N^{-}(t)$. 
To put it in a nutshell, the apparent horizon can emerge by any small perturbation of the scalar field on the boundary far way enough. 
In other words, one can always adjust $r_{\text{min}}$ so that the interval $[r_{\text{min}}, r_{\text{max}}]$ contains the apparent horizon $r_H$, and the change of $r_{\text{min}}$ do not affect the position of the apparent horizon. 
In fact, we are sure that when the outer boundary perturbation is vanished, there must be no apparent horizon, because the (analytic) solution obtained at this time is still a pure AdS solution. 
The apparent horizon $r_H$ will increase as the increasing of amplitude $A$. These results are shown at Tab.\ref{tab1} and Tab.\ref{tab2}. However, this is different from the case without the cosmological constant. Actually, we find no apparent horizon in the case of $\Lambda=0$, when $A$ is not large enough.  
\begin{table}[htbp]
	\centering
	\begin{tabular}{|c|c|c|}
		\hline
		$A$               & $r_H(r_\text{max}=21)$ & $r_H(r_\text{max}=31)$ \\
		\hline
		$1\times10^{-3}$  & 0.7768($*$)            & 1.9525                 \\
		$2\times10^{-3}$  & 1.8000                 & 3.4900                 \\
		$3\times10^{-3}$  & 2.5950                 & 4.7200                 \\
		$4\times10^{-3}$  & 3.2700                 & 5.7925                 \\
		$5\times10^{-3}$  & 3.8750                 & 6.7675                 \\
		$6\times10^{-3}$  & 4.4350                 & 7.6750                 \\
		$7\times10^{-3}$  & 4.9550                 & 8.5225                 \\
		$8\times10^{-3}$  & 5.4500                 & 9.3250                 \\
		$9\times10^{-3}$  & 5.9150                 & 10.0900                \\
		$10\times10^{-3}$ & 6.3650                 & 10.8250                \\
		\hline
	\end{tabular}
	\caption{For $r_\text{min}=1$, the apparent horizon $r_H$ changes under different amplitudes $A$ in this case of type \uppercase\expandafter{\romannumeral1}. We denote that $*$ are calculated at the case $r_\text{min}=0.5$, $r_\text{max}=21$ and $r_\text{min}$ do not impact the apparent horizon $r_H$.}
	\label{tab1}
\end{table}

\begin{table}[htbp]
	\centering
	\begin{tabular}{|c|c|c|}
		\hline
		$A$               & $r_H(r_\text{max}=21)$ & $r_H(r_\text{max}=31)$ \\
		\hline
		$1\times10^{-3}$  & 0.8895($*$)            & 2.1550                 \\
		$2\times10^{-3}$  & 1.9700                 & 3.7975                 \\
		$3\times10^{-3}$  & 2.8100                 & 5.1100                 \\
		$4\times10^{-3}$  & 3.5300                 & 6.2650                 \\
		$5\times10^{-3}$  & 4.1750                 & 7.3150                 \\
		$6\times10^{-3}$  & 4.7700                 & 8.2900                 \\
		$7\times10^{-3}$  & 5.3300                 & 9.2050                 \\
		$8\times10^{-3}$  & 5.8600                 & 10.0750                \\
		$9\times10^{-3}$  & 6.3600                 & 10.9000                \\
		$10\times10^{-3}$ & 6.8450                 & 11.6950                \\
		\hline
	\end{tabular}
	\caption{For $r_\text{min}=1$, the apparent horizon $r_H$ changes under different amplitudes $A$ in this case of type \uppercase\expandafter{\romannumeral2}. We denote that $*$ are calculated at the case $r_\text{min}=0.5$, $r_\text{max}=21$ and $r_\text{min}$ do not impact the apparent horizon $r_H$.}
	\label{tab2}
\end{table}

\section{convergence test}\label{sec_4}
\begin{figure}[htbp]
	\centering
	\includegraphics[width=3.0 in]{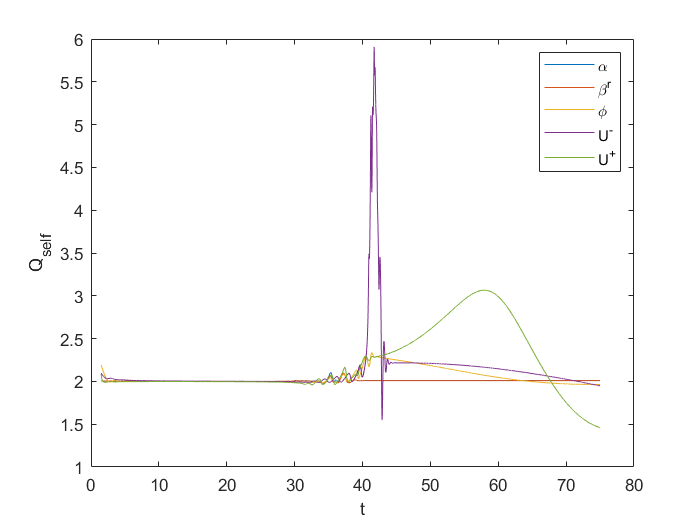}
	\caption{The convergence factors $Q_{\text{self}}$ of five fields $\alpha$, $\beta^r$, $\varphi$, $U^{-}$ and $U^{+}$ change over time, respectively. The outer boundary is at $r_{\text{max}}=31$ with $t_\text{I}=1$, $t_\text{F}=11$, $A=1\times10^{-3}$ in the case of type \uppercase\expandafter{\romannumeral1}. }
	\label{fig_Q}
\end{figure}

Given initial data and boundary data, to calibrate the accuracy of the numerical implementation, we perform simulations at increasing resolution and compare the results. The initial data in these tests are given by Eq.(\ref{initial_data}) and change the location of the outer boundary $r_{\text{max}}$. $N=1001$, $N=2001$, $N=4001$ spatial grid points are carried out in our simulations, respectively. We then estimate the relative error between different resolutions and define the convergence factor as
\begin{eqnarray}
	Q_{\text{self}}=\log_2\Big(\frac{||u_\Delta-u_{\Delta/2}||}{||u_{\Delta/2}-u_{\Delta/4}||}\Big)\, ,\label{Q_factor}
\end{eqnarray}
where $u_{\Delta}$ refers to the numerical solution obtained with resolution $\Delta r=\Delta$. For a grid function $f_i$, the norm defined in Eq.(\ref{Q_factor}) is
\begin{eqnarray}
	||f_i||\equiv\Big[\frac{1}{L}\sum_{i=1}^{N}(f_i)^2\Delta r\Big]^{1/2}\, ,\label{norm}
\end{eqnarray}
where $L=r_{\text{max}}-r_{\text{min}}$. Since the spatial discretization of the radial derivatives is second order and the time integration is also second order, the convergence factor $Q_{\text{self}}$ reduces to $2$ as $\Delta\to0$ in analytical case~\cite{2006Introduction,1910On}. The convergence factors for evolutionary variables are shown in Fig.\ref{fig_Q}. They are approximately $2$, indicating second-order convergence.

\begin{figure}[htbp]
	\centering
	\includegraphics[width=3.0 in]{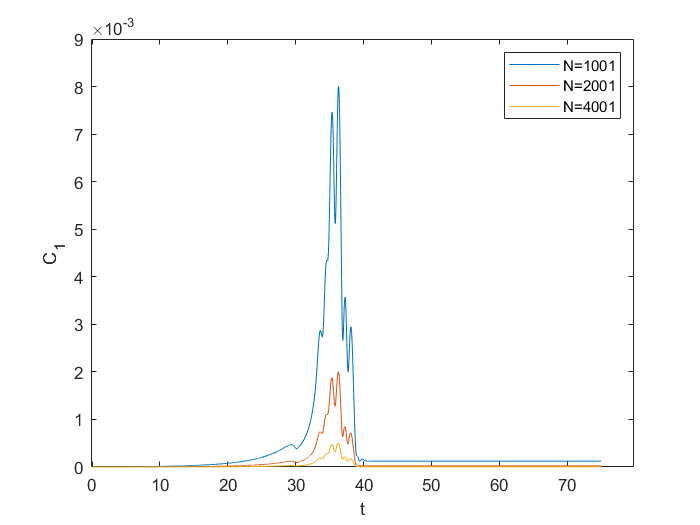}
	\caption{For $r_{\text{min}}=1$, $r_{\text{max}}=31$ with $t_\text{I}=1$, $t_\text{F}=11$, $A=1\times10^{-3}$ in the case of type \uppercase\expandafter{\romannumeral1}. $L_2$ norm of the constraint $\mathcal{C}_1$ with different resolutions}
	\label{constr_1}
\end{figure}

\begin{figure}[htbp]
	\centering
	\includegraphics[width=3.0 in]{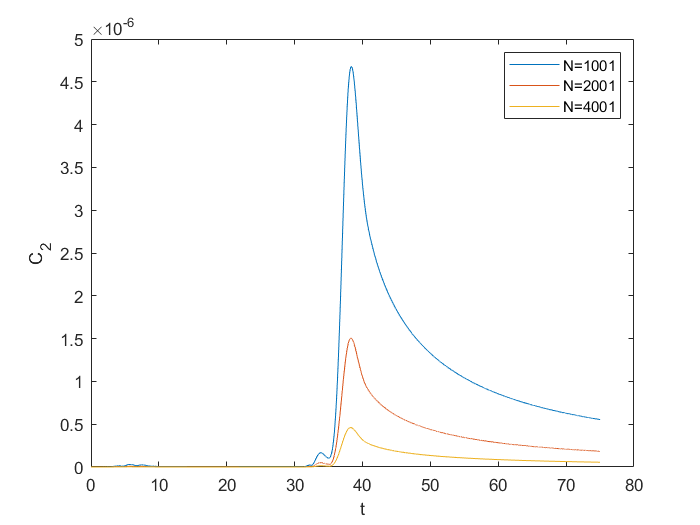}
	\caption{For $r_{\text{min}}=1$, $r_{\text{max}}=31$ with $t_\text{I}=1$, $t_\text{F}=11$, $A=1\times10^{-3}$ in the case of type \uppercase\expandafter{\romannumeral1}. $L_2$ norm of the constraint $\mathcal{C}_2$ with different resolutions}
	\label{constr_2}
\end{figure}

The constraint Eq.(\ref{constraint_1}) is not enforced during the simulation but only used to get the initial data. It is preserved by constructing constraint preserving boundary conditions, although our numerical scheme is a free evolution. The norm of constraint is also defined by Eq.(\ref{norm}). However, some subtle changes need to be explained. When we compute the constraint, we have $L=r(N-1)-r(2)$ with $i=2,\cdots,N-1$ since second order central difference is used to approximate radial derivatives. To tell the truth, the absolute size of the norm of the constraint is meaningless. But, for different resolutions, we will compare with the change of the norm of the constraint with time by keeping the same initial boundary conditions. These results are shown in Fig.\ref{constr_1} and Fig.\ref{constr_2}. From these results, we know the constraints are vanished numerically. This is exactly what we expect.

\section{conclusions and discussion}\label{sec_5}
The ADM and Bondi-Sachs frameworks for the Einstein equations have traditionally been looked upon as two separate approaches for numerical simulations. We have worked out that in the Bondi-Sachs gauge (\ref{advanced_Bondi_Sachs_gauge}), the ADM formulation of Einstein equations with a cosmological constant can be casted into a strongly hyperbolic problem for the lapse and the shift. The ADM formulation in the Bondi-Sachs gauge has one constraint (\ref{constraint_1}) on the initial data for the lapse and shift and one constraint-preserving boundary condition (\ref{boundary_condition}).

In our present work, the assumption of spherical symmetry is kept throughout. We use the constraint preserving formulation to give numerical evidence that the pure AdS spacetime is unstable under slight perturbations. Our simulation is carried out by changing the boundary conditions which is different from the Ref.\cite{Bizon:2011gg} in which a slight perturbation is added to the initial data. In fact, for a pure AdS solution, we find that any small perturbation of the scalar field at the boundary far away enough can  cause the collapse of the pure AdS spacetime. The numerical evidence is provided for the formation of  apparent horizons. The relationship between the perturbation and the position of the apparent horizon has been studied numerically. However, the situation is different from the case with zero cosmological constant. Actually, there is no apparent horizon in the case of $\Lambda=0$, when $A$ is not large enough.  

From the two tables \ref{tab1} and \ref{tab2}, we can see that for the same type of perturbation with the same amplitude, the radius of the apparent horizon will increase as the location of the perturbation staying off. In addition, for the scalar field perturbation placed at the same position, the larger the amplitude of the perturbation, the larger the apparent horizon formed by collapsing finally. Two kinds of perturbation with different shapes express similar conclusions.

\section*{Acknowledgement}
This work was supported in part by the National Natural Science Foundation of China with grants No.12075232, No.11622543, No.11947301, and No.12047502. This work is also supported by the Fundamental Research Funds for the Central Universities under Grant No: WK2030000036.

\appendix
\section{Bianchi identity}\label{Bianchi identity}
In this Appendix, we will give the details for picking out the independent components of $E_{ab}$. At first, when scalar is satisfied with $\nabla_a\nabla^a\varphi=0$, we have the Bianchi identity  which is given by
\begin{eqnarray}
	\nabla^aE_{ab}=-8\pi(\nabla_a\nabla^a\varphi)\nabla_b\varphi=0\, .
\end{eqnarray}
The nontrivial components of $E_{ab}$ are $E_{vv}$, $E_{v\hat{r}}=E_{\hat{r}v}$, $E_{vv}$, $E_{\theta\theta}$, $E_{\phi\phi}$, respectively. It can be proved that $\nabla^\mu E_{\mu\theta}=0$ and $\nabla^\mu E_{\mu\phi}=0$ are trivial under the assumption of spherical symmetry. For $b=v$,
\begin{eqnarray}\label{nabla_E_v}
	\nabla^\mu E_{\mu v}&=&g^{\hat{r}v}\frac{\partial E_{\hat{r}v}}{\partial v}+g^{\hat{r}\hat{r}}\frac{\partial E_{\hat{r}v}}{\partial\hat{r}}+g^{v\hat{r}}\frac{\partial E_{vv}}{\partial\hat{r}}-\Big(3g^{v\hat{r}}\Gamma^{\hat{r}}{}_{\hat{r}v}+g^{v\hat{r}}\Gamma^v{}_{vv}+g^{\hat{r}\hat{r}}\Gamma^{\hat{r}}{}_{\hat{r}\hat{r}}+2g^{\theta\theta}\Gamma^{\hat{r}}{}_{\theta\theta}\Big)E_{\hat{r}v}\nonumber\\
	&&-2g^{\theta\theta}\Gamma^v{}_{\theta\theta}E_{vv}-\Big(g^{v\hat{r}}\Gamma^{\hat{r}}{}_{vv}+g^{\hat{r}\hat{r}}\Gamma^{\hat{r}}{}_{\hat{r}v}\Big)E_{\hat{r}\hat{r}}=0\, .
\end{eqnarray}
For $b=\hat{r}$,
\begin{eqnarray}\label{nabla_E_r}
	\nabla^\mu E_{\mu \hat{r}}&=&g^{v\hat{r}}\frac{\partial E_{v\hat{r}}}{\partial\hat{r}}+g^{\hat{r}v}\frac{\partial E_{\hat{r}\hat{r}}}{\partial v}+g^{\hat{r}\hat{r}}\frac{\partial E_{\hat{r}\hat{r}}}{\partial\hat{r}}-\Big(3g^{v\hat{r}}\Gamma^{\hat{r}}{}_{v\hat{r}}+2g^{\hat{r}\hat{r}}\Gamma^{\hat{r}}{}_{\hat{r}\hat{r}}+2g^{\theta\theta}\Gamma^{\hat{r}}{}_{\theta\theta}\Big)E_{\hat{r}\hat{r}}\nonumber\\
	&&-\Big(g^{v\hat{r}}\Gamma^{\hat{r}}{}_{\hat{r}\hat{r}}+2g^{\theta\theta}\Gamma^v{}_{\theta\theta}\Big)E_{v\hat{r}}-2g^{\theta\theta}\Gamma^{\theta}{}_{\theta\hat{r}}E_{\theta\theta}=0\, .
\end{eqnarray}
From Eq.(\ref{nabla_E_r}), when $E_{v\hat{r}}=E_{\hat{r}\hat{r}}=0$, we have $E_{\theta\theta}=0$. Then one gets $E_{\phi\phi}=0$, since $E_{\phi\phi}=\sin^2\theta E_{\theta\theta}$. From Eq.(\ref{nabla_E_v}), when $E_{v\hat{r}}=E_{\hat{r}\hat{r}}=0$, we obtain
\begin{eqnarray}\label{E_v^r_1}
	g^{v\hat{r}}\frac{\partial E_{vv}}{\partial \hat{r}}-2g^{\theta\theta}\Gamma^v{}_{\theta\theta}E_{vv}=0\, .
\end{eqnarray}
This equation is equivalent to
\begin{eqnarray}\label{E_v^r_2}
	\frac{\partial(E_{vv}g^{v\hat{r}})}{\partial\hat{r}}-\frac{\partial g^{v\hat{r}}}{\partial\hat{r}}E_{vv}-2g^{\theta\theta}\Gamma^v{}_{\theta\theta}E_{vv}=0\, .
\end{eqnarray}
If $E_{v\hat{r}}=0$, we have $E_v{}^{\hat{r}}=E_{v\mu}g^{\mu\hat{r}}=E_{vv}g^{v\hat{r}}$. For $g^{v\hat{r}}=e^{-2\beta}$, we have $\partial g^{v\hat{r}}/\partial\hat{r}=-2g^{v\hat{r}}\partial\beta/\partial\hat{r}$. Therefore, Eq.(\ref{E_v^r_2}) becomes
\begin{eqnarray}\label{E_v^r_3}
	\frac{\partial E_v{}^{\hat{r}}}{\partial\hat{r}}+2\Big(\frac{\partial\beta}{\partial \hat{r}}+\frac{1}{\hat{r}}\Big)E_v{}^{\hat{r}}=0\, .
\end{eqnarray}
Solving this equation (\ref{E_v^r_3}), one gets~\cite{Christodoulou:1986zr}
\begin{eqnarray}
	E_v{}^{\hat{r}}(v,\hat{r})=\frac{\hat{r}^2_0}{\hat{r}^2}\exp2\Big[\beta(v,\hat{r}_0)-\beta(v,\hat{r})\Big]E_v{}^{\hat{r}}(v,\hat{r}_0)\, .
\end{eqnarray}
Therefore, the independent components of $E_{ab}$ are $E_{\hat{r}\hat{r}}$ and $E_{v\hat{r}}$. That means evolution equations are
\begin{eqnarray}\label{Bondi_EOM_1}
	&&E_{\hat{r}\hat{r}}=0\, ,\nonumber\\
	&&E_{v\hat{r}}=0\, ,\nonumber\\
	&&\nabla_a\nabla^a\varphi=0\, .
\end{eqnarray}
According to the solution of Eq.(\ref{E_v^r_3}),
\begin{eqnarray}
	E_v{}^{\hat{r}}=0
\end{eqnarray}
is recommended to be a constraint equation.
\section{calculation of basic quantity}
In this Appendix, we give results of some basic quantities which we have used in the process of deriving evolution equation. The non-vanished Christoffel symbols are given as follows,
\begin{eqnarray}\label{Bondi_Christoffel}
	\Gamma^v{}_{vv}&=&\frac{1}{2\hat{r}^2}\Big(-V+2\hat{r}V\beta_{,\hat{r}}+\hat{r}V_{,\hat{r}}+4\hat{r}^2\beta_{,v}\Big)\, ,\nonumber\\
	\Gamma^v{}_{\theta\theta}&=&-\hat{r}e^{-2\beta}\, ,\nonumber\\
	\Gamma^v{}_{\phi\phi}&=&-\hat{r}e^{-2\beta}\sin^2\theta\, ,\nonumber\\
	\Gamma^{\hat{r}}{}_{vv}&=&\frac{1}{2\hat{r}^3}\Big(-V^2+2\hat{r}V^2\beta_{,\hat{r}}-\hat{r}^2V_{,v}+\hat{r}VV_{,\hat{r}}+2\hat{r}^2V\beta_{,v}\Big)\, ,\nonumber\\
	\Gamma^{\hat{r}}{}_{v\hat{r}}&=&\Gamma^{\hat{r}}{}_{\hat{r}v}=\frac{1}{2\hat{r}^2}\Big(V-\hat{r}V_{,\hat{r}}-2\hat{r}V\beta_{,\hat{r}}\Big)\, ,\nonumber\\
	\Gamma^{\hat{r}}{}_{\hat{r}\hat{r}}&=&2\beta_{,\hat{r}}\, ,\nonumber\\
	\Gamma^{\hat{r}}{}_{\theta\theta}&=&-Ve^{-2\beta}\, ,\nonumber\\
	\Gamma^{\hat{r}}{}_{\phi\phi}&=&-Ve^{-2\beta}\sin^2\theta\, ,\nonumber\\
	\Gamma^{\theta}{}_{\hat{r}\theta}&=&\Gamma^{\theta}{}_{\theta\hat{r}}=\frac{1}{\hat{r}}\, ,\nonumber\\
	\Gamma^{\theta}{}_{\phi\phi}&=&-\cos\theta\sin\theta\, ,\nonumber\\
	\Gamma^{\phi}{}_{\hat{r}\phi}&=&\Gamma^{\phi}{}_{\phi\hat{r}}=\frac{1}{\hat{r}}\, ,\nonumber\\
	\Gamma^{\phi}{}_{\theta\phi}&=&\Gamma^{\phi}{}_{\phi\theta}=\cot\theta\, .
\end{eqnarray}
The components of the Einstein tensor are
\begin{eqnarray}\label{Bondi_Einstein}
	G_{vv}&=&\frac{1}{\hat{r}^3}\Big(2V^2\beta_{,\hat{r}}-\hat{r}V_{,v}+Ve^{2\beta}-VV_{,\hat{r}}+2\hat{r}V\beta_{,v}\Big)\, ,\nonumber\\
	G_{v\hat{r}}&=&G_{\hat{r}v}=\frac{1}{\hat{r}^2}\Big(-2V\beta_{,\hat{r}}+V_{,\hat{r}}-e^{2\beta}\Big)\, ,\nonumber\\
	G_{\hat{r}\hat{r}}&=&\frac{4}{\hat{r}}\beta_{,\hat{r}}\, ,\nonumber\\
	G_{\theta\theta}&=&\frac{e^{-2\beta}}{2}\Big(-2V\beta_{,\hat{r}}+2\hat{r}V\beta_{,\hat{r}\hat{r}}+2\hat{r}V_{,\hat{r}}\beta_{,\hat{r}}+\hat{r}V_{,\hat{r}\hat{r}}+4\hat{r}^2\beta_{,v\hat{r}}\Big)\, ,\nonumber\\
	G_{\phi\phi}&=&\frac{e^{-2\beta}}{2}\Big(-2V\beta_{,\hat{r}}+2\hat{r}V\beta_{,\hat{r}\hat{r}}+2\hat{r}V_{,\hat{r}}\beta_{,\hat{r}}+\hat{r}V_{,\hat{r}\hat{r}}+4\hat{r}^2\beta_{,v\hat{r}}\Big)\sin^2\theta\, ,
\end{eqnarray}
and
\begin{eqnarray}
	G_{v}{}^{\hat{r}}=\frac{e^{-2\beta}}{\hat{r}^2}\Big(2V\beta_{,v}-V_{,v}\Big)\, .
\end{eqnarray}
The components of $\nabla_a\nabla_b\varphi$ are
\begin{eqnarray}\label{Bondi_nabla_nabla_varphi}
	\nabla_v\nabla_v\varphi&=&\varphi_{,vv}-\frac{1}{2\hat{r}^2}\Big(-V+2\hat{r}V\beta_{,\hat{r}}+\hat{r}V_{,\hat{r}}+4\hat{r}^2\beta_{,v}\Big)\varphi_{,v}\nonumber\\
	&&-\frac{1}{2\hat{r}^3}\Big(-V^2+2\hat{r}V^2\beta_{,\hat{r}}-\hat{r}^2V_{,v}+\hat{r}VV_{,\hat{r}}+2\hat{r}^2V\beta_{,v}\Big)\varphi_{,\hat{r}}\, ,\nonumber\\
	\nabla_v\nabla_{\hat{r}}\varphi&=&\nabla_{\hat{r}}\nabla_v\varphi=\varphi_{,v\hat{r}}-\frac{1}{2\hat{r}^2}\Big(V-\hat{r}V_{,\hat{r}}-2\hat{r}V\beta_{,\hat{r}}\Big)\varphi_{,\hat{r}}\, ,\nonumber\\
	\nabla_{\hat{r}}\nabla_{\hat{r}}\varphi&=&\varphi_{,\hat{r}\hat{r}}-2\beta_{,\hat{r}}\varphi_{,\hat{r}}\, ,\nonumber\\
	\nabla_{\theta}\nabla_{\theta}\varphi&=&Ve^{-2\beta}\varphi_{,\hat{r}}+\hat{r}e^{-2\beta}\varphi_{,v}\, ,\nonumber\\
	\nabla_{\phi}\nabla_{\phi}\varphi&=&\Big(Ve^{-2\beta}\varphi_{,\hat{r}}+\hat{r}e^{-2\beta}\varphi_{,v}\Big)\sin^2\theta\, .
\end{eqnarray}
The expression of $\nabla_a\nabla^a\varphi$ is given by
\begin{eqnarray}
	\nabla_a\nabla^a\varphi=e^{-2\beta}\Big[2\varphi_{,v\hat{r}}+\frac{V}{\hat{r}}\varphi_{,\hat{r}\hat{r}}+\Big(\frac{V_{,\hat{r}}}{\hat{r}}+\frac{V}{\hat{r}^2}\Big)\varphi_{,\hat{r}}+\frac{2}{\hat{r}}\varphi_{,v}\Big]\, .
\end{eqnarray}
The expression of $\nabla_a\varphi\nabla^a\varphi$ is given by
\begin{eqnarray}
	\nabla_a\varphi\nabla^a\varphi=\frac{V}{\hat{r}}e^{-2\beta}\varphi_{,\hat{r}}\varphi_{,\hat{r}}+2e^{-2\beta}\varphi_{,\hat{r}}\varphi_{,v}\, .
\end{eqnarray}
\bibliography{reference}{}
\bibliographystyle{apsrev4-1}

\end{document}